\documentclass[fleqn,10pt]{wlscirep}
\usepackage{verbatim}
\usepackage{xcolor}
\usepackage{url}

\title{A Parametric Study of Radiative Dipole Body\\ Array Coil for 7 Tesla MRI}

\author[1,*]{Anna A. Hurshkainen}
\author[2]{Bart Steensma}
\author[1]{Stanislav B. Glybovski}
\author[2]{Ingmar J. Voogt}
\author[1]{\\Irina V. Melchakova}
\author[1]{Pavel A. Belov}
\author[2]{Cornelis A.T. van den Berg}
\author[2]{\\Alexander J.E. Raaijmakers}
\affil[1]{Faculty of Physics and Engineering, ITMO University, 197101 St. Petersburg, Russia}
\affil[2]{University Medical Center Utrecht, 3584 CX, Netherlands}

\affil[*]{a.hurshkainen@metalab.ifmo.ru}

\begin{abstract}
In this contribution we present numerical and experimental results of a parametric quantitative study of radiative dipole antennas in a phased array configuration for efficient body magnetic resonance imaging at 7T via parallel transmission. For magnetic resonance imaging (MRI) at ultrahigh fields (7T and higher) dipole antennas are commonly used in phased arrays, particularly for body imaging targets. This study reveals the effects of dipole positioning in the array (elevation of dipoles above the subject and inter-dipole spacing) on their mutual coupling, $B_1^{+}$ per $P_{acc}$ and $B_1^{+}$ per maximum local SAR efficiencies as well as the RF-shimming capability. The numerical and experimental results are obtained and compared for a homogeneous phantom as well as for a real human models confirmed by \textit{in-vivo} experiments. 
\end{abstract}

\begin{document}

\maketitle

\thispagestyle{empty}

\section*{Introduction}

MRI requires the use of radiofrequency (RF) antennas to excite selected atomic nuclei in a subject and to detect echo signals from relaxing nuclei. For historical reasons, these antennas are referred to as RF-coils. The latter may be transmit coils and/or receive coils. Clinical high field MRI systems with a permanent magnetic field strength $B_0$ from 1.5 to 3.0 T,  use a combination of a bird-cage body coil \cite{HAYES1985622} in transmit and a local multi-channel coil array in receive \cite{surfaceloop}. For ultrahigh field MRI with a $B_0$ field strength of 7T (or higher) the Larmor frequency of protons reaches 298 MHz. The higher is $B_0$ the better are SNR and contrast of images for the same scan duration \cite{uugurbil2003ultrahigh,zwanenburg2008mr,kollia2009first}. However, at higher Larmor frequencies the wavelength in body tissues becomes comparable to human torso dimensions. As a result, the wave propagation and interference effects distort the distribution of the RF magnetic field $B_1^{+}$ created by a transmit coil. Making the conventional approach with a bird-cage body coil not anymore feasible \cite{Vaughan2009}. In particular, considerable inhomogeneity artifacts occur locally reducing the image quality.

The most efficient solution to this problem is the parallel transmission approach \cite{JMRI:JMRI6,IBRAHIM2000835,MRM:MRM10353,MRM:MRM20011}, in which the RF field in the region of interest (ROI) can be optimized by manipulating with phases and sometimes magnitudes of RF-signals applied at inputs of antenna array elements to maximize locally the $B_1^{+}$ for a given transmitter power or to reach the most uniform $B_1^{+}$ distribution over the whole ROI.

Ultrahigh field body imaging requires a multi-transmit system with transmit/receive phased arrays of antennas arranged at the body surface (so-called \textit{surface coil arrays}). A wide variety of array designs have been proposed consisting of microstrip antenna elements \cite{Shim_prost, orzada2009flexible, Cardiac7T}, loop \cite{Cardiac16ch_7T} or dipole antennas \cite{Raaijmakers_dipole, Winter_Hyperthermia, wiggins2013mixing, Fractionated}.  
Dipole antennas are particularly useful for body imaging at frequencies corresponding to the ultrahigh fields and/or deeply located imaging targets \cite{Ideal_current_pattern, Dip_loop_comp}. The earliest use of dipole antennas for transceive array purposes was the \textit{single-side adapted dipole antenna} \cite{Raaijmakers_dipole}. This antenna uses a high-permittivity ceramic substrate which reduces its electrical length and the local specific absorption rate (SAR) on the surface of the subject (a human body) for the same $B_1^{+}$  at depth. This property reliefs to some extent limitations associated with power absorption by the body tissues.  This dipole design was adapted by Winter et al. into what has been called  \textit{Bow-tie} antennas \cite{Winter_Hyperthermia} where the width of the antenna legs was diverging to realize a bow-tie shape. Wiggins et al. presented a design on the use of dipole antennas for head imaging with a meandering bifurcation at the endings of the antenna legs \cite{wiggins2012electric}. The same group has recently presented a wide variety of dipole-loop hybrids such as the \textit{loopole} \cite{lakshmanan2014loopole} and the circular dipole \cite{lakshmanan2014circular}. Also a modified dipole antenna design was presented with increased directivity due to phase matching of the full antenna length \cite{Curved_dipole}. Further developments in the use of dipole antennas include the use of dipole antennas for spine \cite{Spine_dip} and brain imaging applications \cite{chen20147t}. Dipole antennas have also been used to generate so-called \textit{dark modes} that have no effect on image intensity but reduce SAR levels by destructive interference of the electric fields \cite{Dark_modes}.

A particularly attractive design has recently been presented by Raaijmakers et al. \cite{Fractionated} where the legs of the dipole have been split into segments with inter-segment meanders (inductors). The design can be used easily in combination with two additional loop coils for receive per antenna, resulting in a 8-channel transmit/24-channel receive array. With this setup, superior SNR over 3T was recently demonstrated for prostate imaging at 7T \cite{Ingmar_comb}. Although the intrinsic gain in SNR was established, the comparison did not yet take into account the additional restrictions at 7T caused by local SAR limitations. Using local transceive arrays of dipole antennas, local SAR is the limiting factor for imaging speed at ultrahigh field strengths. Additional reduction of local SAR levels seems necessary to be able to outperform 3T in multislice TSE imaging, which is the clinical work horse for diagnostic imaging of prostate cancer.

In principle, SAR level can be reduced by increasing the distance from the antenna to the tissue, because the divergent RF energy coming from the antenna is smeared out over a larger area resulting in lower local energy deposition. This trend still needs a revision with careful consideration of the local SAR distribution as a function of the distance. This would help to optimize the practical configuration of body arrays consisting of fractionated dipole antennas.
On the other hand, one could also assume that the increased spacing from an antenna to the body tissues will reduce the antenna efficiency meaning that the $B_1^{+}$ per unit accepted power will degrade. 
In addition, the reduced loading of the antenna is expected to result in a higher inter-element coupling in the array configuration, which would lead to even lower efficiencies and difficulties in independent tuning and matching of the array elements. Finally, higher Q-factors are expected to reduce the element bandwidth making it more susceptible to loading variations.

This study tests the above assumptions by parametric simulations and bench measurements with the variation of the dipole-subject distance. The study is divided in two steps as follows. On the first step a single dipole array element was investigated being placed at different distances to a subject. Moreover, two identical dipole elements were studied both placed at the same distance to a subject. The aim of this step was to qualitatively estimate the main dipole element's characteristics (i.e. the $B_1^{+}$ per unit square root of accepted power efficiency, maximum local SAR per unit accepted power, inter-subject stability of matching and inter-element coupling) for different inter-element and antenna-subject distances. Here numerical simulations and measurements were made on phantoms.
On the second step, the whole-array setup was studied versus the antenna-subject distance keeping the same number of elements (eight) and the same inter-element distance to parametrically study how variation of this distance affects $B_1^{+}$ shimming capabilities both by simulations on voxel body model and \textit{in-vivo} measurements. The eight channel configuration was chosen to be studied in this paper due to sufficiently low inter-element coupling for the whole range of studied antenna-subject distances and to compare with the original setup from the previous paper \cite{Fractionated}. Finally, an array setup adapted using the parametric results was tested for prostate imaging at 7T in comparison to the original setup. 

\section*{Methods}

This study will investigate the impact of an increased antenna-subject spacing on SAR levels, transmit efficiency/receive sensitivity, inter-element coupling and the robustness towards loading variations for fractionated dipole antennas. The comparison is made to the original setup \cite{Fractionated}, where the antenna-subject spacing is 18 mm. At the first step the simulation and measurement studies of both the inter-element and the antenna-subject distances conducted for only two dipoles were made to qualitatively reveal the reachable values of the above characteristics and show that the choice of 8 array elements is indeed reasonable in terms of inter-element coupling for the observed range of the inter-element spacing. The aim of the second step was to observe qualitatively the ranges of main transmit characteristics and inter-element coupling versus the antenna-subject distance for the 8-element configuration. Finally, based on the outcomes of the corresponding parametric study, $B_1^{+}$ mapping and prostate imaging were performed using the parallel transmission with equal signal magnitudes at all 8 array channels. 

\subsection*{Study of single array elements}

\subparagraph*{SAR levels and transmit efficiency}

The impact of the antenna-subject spacing on SAR levels and transmit efficiency has been calculated at the Larmor frequency of 298 MHz by numerical simulations using the finite element method in Ansys HFSS commercial software. The simulation model contained a single fractionated dipole (a pair of meandered copper traces on a FR4 1.5-mm thick substrate fixed on a rectangular polycarbonate placeholder), a phantom which is filled with the uniform dielectric material with the permittivity of 78 and the electric conductivity of 0.47 S/m. The phantom had dimensions of $650\times380 \times270$ mm.

\begin{figure}[t]
\center
\includegraphics[width=0.9\linewidth]{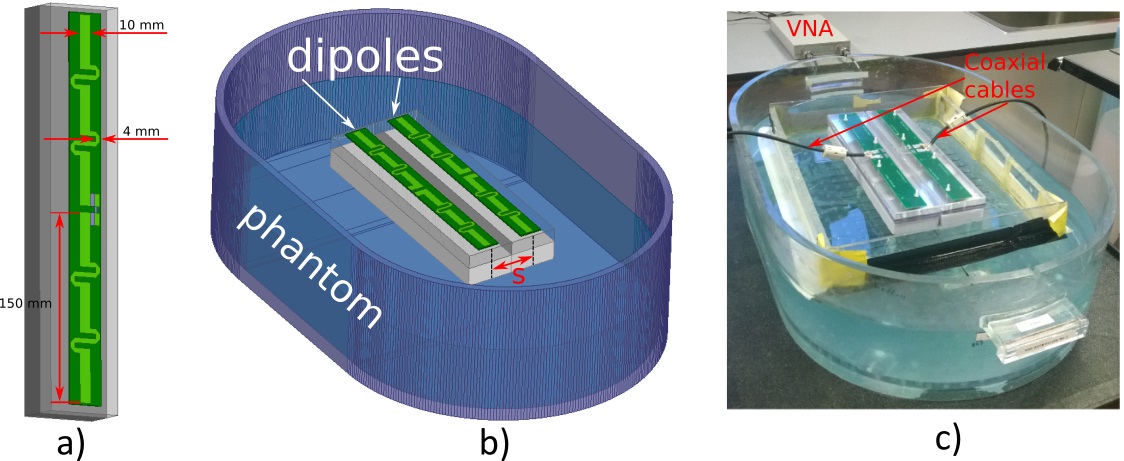}
\caption{ a) Fractionated dipole antenna design; b) simulation model in Ansys HFSS for the inter-element coupling evaluation of two dipole antennas over a homogeneous phantom (images a,b were created using Ansys HFSS); c) setup for on-bench inter-element coupling measurements of two dipole antennas over a homogeneous liquid phantom}
\label{setup}
\end{figure}

The conductors of the meandered dipoles were specified in the simulation as perfectly conducting (PEC) and had the shape and dimensions given in Fig. \ref{setup}a. To ensure validity of using PEC conductors a comparative simulation was performed where all conductors were implemented as copper for H = 73 which is the least loaded case, and hence, then the intrinsic losses of a coil give a highest impact at the coil performance. The obtained results shown the 2.4\% reduction of the $B_1^{+}$ at the surface of the phantom for copper coil with respect to PEC. Since this is the worst case in terms of the impact of antenna intrinsic losses, for other H this discrepancy is even lower, which allows with a sufficient accuracy use PEC antenna in simulations. The substrate of the dipoles was implemented as dielectric having dimensions of $30\times1.5\times308$ mm,  permittivity of 4.4 and the dielectric loss tangent of 0.02. The polycarbonate placeholder had the dimensions of $64\times320\times18$ mm, with a permittivity of 3.4 and a dielectric loss tangent of 0.001. In the original array setup \cite{Fractionated} the spacing between the subject and the conductors of the dipole was almost equal to the 18 mm thickness of the placeholder (the placeholder was touching the subject's surface). In order to increase the antenna-subject spacing in comparison to the original setup, an additional foam layer of thickness $h$ was placed between the subject and the placeholder, which has been represented in the simulations by an air-filled volume with a permittivity of 1.0 and zero conductivity (approximately corresponds to the material parameters of a foam substrate at 298 MHz). Simulations were done for three values of the additional foam spacing $h$. The original value of $h=0$ (placeholder touches the phantom and the total elevation $H$ of the dipole above the phantom is equal to the 18 mm thickness of the placeholder). Three values of the additional spacing: $h=25$ mm (total elevation $H$ over the phantom 43 mm), $h=40$ mm ($H=58$ mm)  and $h=55$ mm ($H=73$ mm). The environment of 7T scanner was taken into account in simulations by means of a metallic bore with the diameter of 60 cm and the length of 1 m. Each simulation was done for 1W of accepted power with the corresponding normalization of the $B_1^{+}$ and local SAR distributions. From these quantities the SAR efficiency was calculated as the $B_1^{+}$ level at a particular depth of the region of interest (ROI) in the phantom divided by the square root of the maximum 10g averaged local SAR value (which was observed typically at the surface of the phantom right below the dipole's center). 

\subparagraph*{Inter-element coupling}

Since inter-element coupling is sometimes reported as difficult to simulate correctly, the impact of antenna-subject spacing on inter-element coupling has been determined by measurements, supported by numerical simulations. 
Dipoles were manufactured using printed circuit board technology (PCB) on FR4 laminate material (Eurocircuits N.V., Mechelen, Belgium). PCB was attached to the polycarbonate placeholder. Additional foam layer was realized using lossless dielectric material (Flamex Basic) with a permittivity of 1. For the experimental study two similar fractionated dipole antennas (Fig. \ref{setup}c) were positioned on a ASTM phantom filled with saline water (2.5 g/L). Measurements of the coupling coefficient $S_{12}$ were performed for a range of inter-element distances $s$ (64 to 124 mm with 10 mm increments). 

These measurements were repeated for a range of antenna-phantom spacing values $H$ (43, 58, 73 mm). S-parameters of the two ports in the two dipoles were measured by means of a vector network analyzer (VNA) Planar TR1300/1 without matching circuitry (direct connections of coax cables to the antennas). Actual coupling coefficient of two antennas was estimated using the formula $S_{12,\text{norm}} = |S_{12}|/{\sqrt{1-|S_{11}|^2}}$ (assuming that S-matrix of the setup with two identical antennas is symmetric). The coupling coefficient vs. inter-element distance was obtained for each of the investigated antenna-phantom spacing values.

To verify the experimental data, the same setup was simulated using Ansys HFSS. The simulation model (Fig. \ref{setup}b) contained the ASTM phantom filled with saline water ($\varepsilon$ = 81, $\sigma$  = 0.47 S/m). S-parameters were calculated for the same antenna-phantom spacing values ($H=43$, 58 and 73 mm) as well as for different inter-element distances ($s=$ 64 – 124 mm). Simulations were compared with measurements on the same graph. 

\subparagraph*{Robustness towards loading variation}

As results will show, the preceding parameter studies favor the largest antenna-subject spacing due to optimal $B_1^{+}/\sqrt{\text{SAR}_{\text{max}}}$ ratio in the depth of a subject. However, our practical experience tells us that with such large spacings the antenna becomes increasingly difficult to match to a 50-Ohm coaxial cable. Particularly obtaining a good match for a wide range of subjects with one fixed matching network is difficult. To investigate this additional downside of increased antenna-subject spacing, the robustness of matching towards loading variations was investigated. Fractionated dipole antennas with all the above mentioned spacing values $H=18$, 43, 58 and 73 mm have been matched on a pelvic shaped phantom with tissue-equivalent dielectric properties (ethylene glycol with 34 g/l NaCl resulting in $\varepsilon$ = 34 and $\sigma$ = 0.4 S/m). The $S_{11}$ curve representing the reflection coefficient in the cable from the antenna port over the frequency range from 250 to 350 MHz has been measured for the phantom and for a set of four volunteers (age 25 – 34, BMI 20 – 25). The presence of the RF shield of the scanner was taken into account by performing all measurements inside a dummy-bore (60 cm diameter, 120 cm length glass fiber cylinder covered with copper tape from the inside).

After the $S_{11}$ curves were obtained for all volunteers, their shapes are compared for different investigated antenna-subject spacing values. 

\subsection*{Study of 8-channel array configuration}

\subparagraph*{Numerical simulations of 8-channel array}

From the parameter study performed using a homogeneous phantom, an optimal antenna-subject spacing of $H=43$ mm has been determined (18 mm polycarbonate + 25 mm foam). This antenna-subject spacing is optimal in terms of $B_1^{+}/\sqrt{\text{SAR}_{\text{max}}}$ ratio in a combination with low sensitivity of the accepted power in each channel to loading variations. This value has been used to realize a 8-element dipole array with increased antenna-subject spacing, in comparison to the original array \cite{Fractionated}, but having the same number of elements. This array as well as a setup with $H=18$ and $H=58$ mm has been simulated using the commercial software Sim4Life \cite{Sim4life} on the voxel human body model “Duke” from Virtual Family \cite{Virt_family}. This way the effect of the antenna elevation on the maximum local SAR can be estimated correctly, in contrast to simulations with a homogeneous phantom. The environment of a 7T MR scanner was represented in the simulations as a cylindrical copper RF-shield with the diameter of 60 cm and the length of 2 m. $B_1^{+}/{\sqrt{P_{acc}}}$ and SAR$/P_{acc}$ maps were calculated after the phase shimming was applied. The phase shimming for three array configurations ($H = 18$, 43, 58 mm) was done by reaching max$(B_1^{+}/{\sqrt{P_{acc}}})$ in the prostate, using a fixed input power (equal drive magnitudes) for each channel and determining the optimal phase settings. The optimal drive phases were calculated based on a simplex minimization in Matlab ("fminsearch" function, Matlab Mathworks, Natick, USA) and are presented in Table \ref{TabShim}. Standard deviations of the $(B_1^{+}/{\sqrt{P_{acc}}})$ in the prostate were: 0.17 uT$/{\sqrt{W}}$ for H = 18 mm, 0.11 uT$/{\sqrt{W}}$ for H = 43 mm and 0.10 uT$/{\sqrt{W}}$ for H = 58 mm. The reported $(B_1^{+}/{\sqrt{P_{acc}}})$ values are averaged over the whole prostate, while the reported SAR values are the maximum SAR values in the body. By doing so, the complete volume of interest is taken into account in the analysis.
\begin{table}
\caption{RF-shimming phase setting of the antenna array elements for 3 setups with different antenna-subject spacing $H$}
\centering
\begin{tabular}{|c|c|c|c|c|c|c|c|c|} 
    \cline{2-9}
    \multicolumn{1}{c|}{} & \multicolumn{8}{c|}{Number of antenna array element}\\ 
    \hline
    $H$, mm & 1 & 2 & 3 & 4 & 5 & 6 & 7 & 8\\
    \hline
    18 & 0 & -62.8 & 143.2 & 83.8 & -123.7 & 103.8 & 150.4 & -50.2\\
    43 & 0 & -48.5 & 154.0 & 64.5 & -145.1 & 114.6 & 139.7 & -84.3\\
    58 & 0 & -44.8 & 153.9 & 58.1 & -152.3 & 114.5 & 134.2 & -98.5\\
    \hline                  
\end{tabular}
\label{TabShim}
\end{table}

\subparagraph*{$B_1^{+}$ mapping and volunteer imaging}

Two arrays with $H=18$ and 43 mm were experimentally compared by $B_1^{+}$ mapping on volunteers (all of them filled out a written form of informed consent). All \textit{in-vivo} studies were approved according to general approval for development studies from the ethical advisory board (Medisch Ethische Toetsingsingscommissie or METc) and were performed in accordance with the relevant guidelines and regulations. Measurements were performed on the Phillips Achieva 7T platform. $B_1^{+}$ maps were acquired by a DREAM sequence \cite{MRM:MRM24158} (TE/TR = 1.97/7.0 ms, FA = 10, STE angle = 60, voxel size 5x5x10 mm$^3$, FOV 400x320x10 mm, scan duration 0.47 s) on a three healthy volunteers (BMI = 20.9, 24.6 26.6, age 26, 27, 38 years).

Subsequently, arrays with $H=18$ and 43 mm has been used for prostate imaging on a healthy volunteer (BMI = 25, age 34). All the elements of the array were matched using similar symmetric LC matching circuits composed of two series inductances. The in-situ coupling matrix has been measured using directional couplers of the scanner. $T_{2}$-weighed images (TR/TE=2500/90 ms, 0.5x0.5x3 mm3, TSE-factor 9) have been performed using both arrays on the same volunteer.

\section*{Results}

\subsection*{Study of single array elements}

\begin{figure}[ht]
\center
\includegraphics[width=1\linewidth]{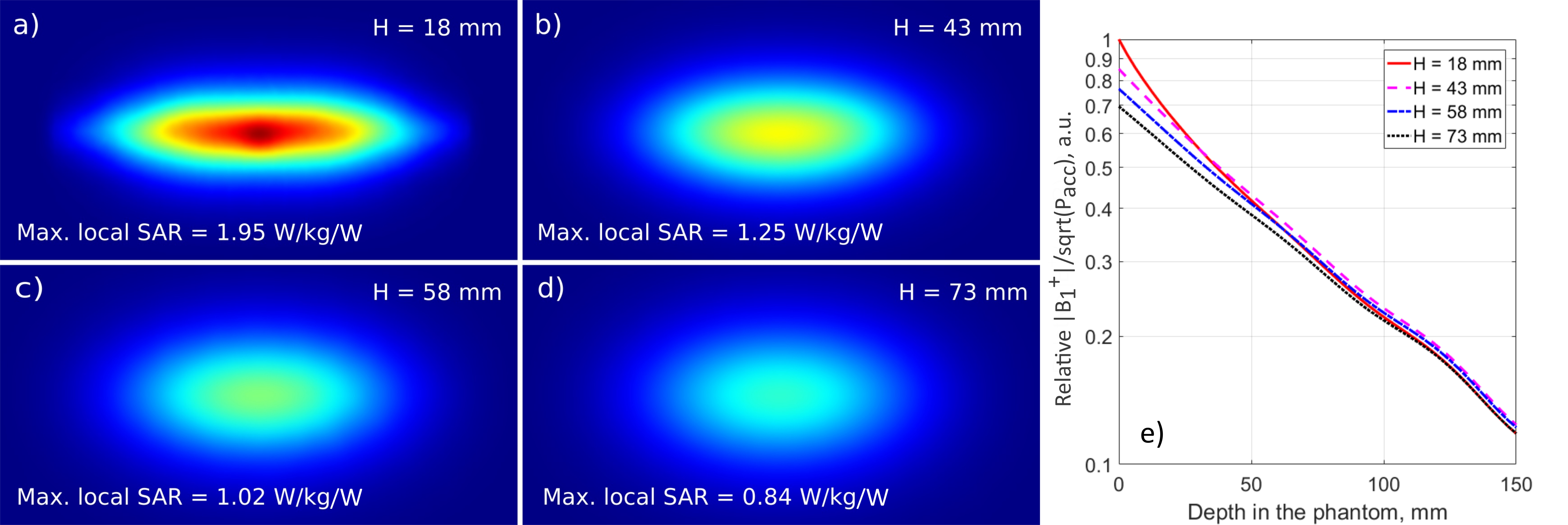}
\caption{Simulation of a single dipole antenna over a homogeneous phantom: (a-d) local SAR$/{P_{acc}}$ of a single dipole antenna in top coronal plane of the phantom for various antenna-subject spacing $H$ ( a) $H$=18mm; b) $H$=43mm; c) $H$=58mm; d) $H$=73mm); e) relative $B_1^{+}/{\sqrt{P_{acc}}}$ profiles created by a single dipole antenna with different antenna-subject spacing ($H$)}
\label{SAR}
\end{figure}

\begin{figure}[t]
\center
\includegraphics[width=1\linewidth]{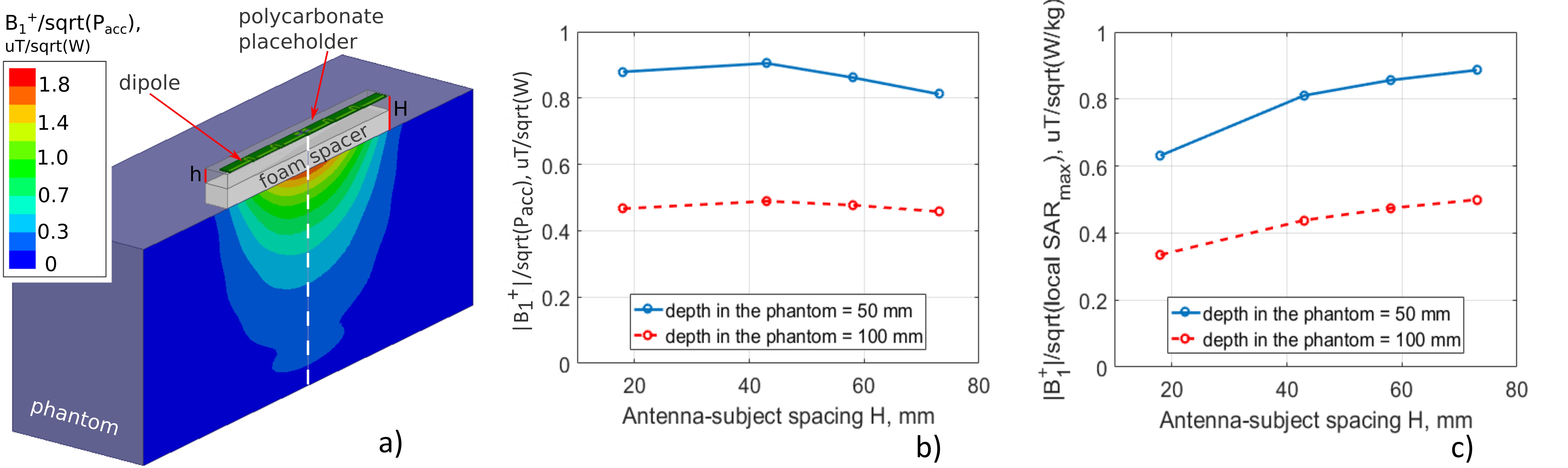}
\caption{Simulation of a single dipole antenna over a homogeneous phantom: a) simulation model (the color map demonstrates $B_1^{+}/{\sqrt{P_{acc}}}$ distribution in the cross-section of the phantom for $H=43$ mm); b) $B_1^{+}/{\sqrt{P_{acc}}}$ values at 50 and 100 mm depth inside the phantom; c) $B_1^{+}/\sqrt{\text{SAR}_{\text{max}}}$ values at 50 and 100 mm depth inside the phantom}
\label{in_depth}
\end{figure}

\subparagraph*{SAR levels and transmit efficiency}
Local SAR$/P_{acc}$ distributions and $B_1^{+}/{\sqrt{P_{acc}}}$ distributions for fractionated dipole antennas for various antenna-phantom spacings have been simulated. The resulting local SAR$/P_{acc}$ distributions in the top coronal plane of the phantom directly beneath the dipole are indicated in Fig. \ref{SAR}a-d.
With additional spacing, SAR levels are reduced as was expected. The maximum local SAR value for the original case ($H=18$ mm) and for the dipole with the additional foam spacer ($H=43$ mm) was 1.95 and 1.25 $\frac{W}{kg}/W$ respectively. This indicates that the maximum local SAR in the phantom values can be reduced by 36\% by increasing the antenna-subject spacing by 25 mm. Further increase of the antenna-subject spacing realizes even more reduction of the maximum local SAR. For the dipoles with $H=58$ and 73 mm maximum local SAR values are 1.02 and 0.84 respectively (48 and 57\% reduction).
The corresponding $B_1^{+}/{\sqrt{P_{acc}}}$ profiles are shown in Fig. \ref{SAR}e. 
These depth profiles were plotted versus the axis going through the center of the dipole, normally to the subject's top surface (depicted in Fig.\ref{in_depth}a by the dashed white line). Clearly, larger antenna-subject spacing values result in lower $B_1^{+}/{\sqrt{P_{acc}}}$ at the surface. At larger depths (beyond 30 mm) the $B_1^{+}/{\sqrt{P_{acc}}}$ with 18 mm spacing is outperformed by the 43 mm spacing curve. This graph continuously decreases more steeply than the others so at larger depths also the other graphs outperform the 18 mm curve. All the other curves are more or less parallel: larger spacing results in reduced $B_1^{+}/{\sqrt{P_{acc}}}$ magnitude regardless of the depth of interest. However, the differences between the curves for any depth are small. Fig. \ref{in_depth}b shows the $B_1^{+}/{\sqrt{P_{acc}}}$ value as a function of antenna-phantom spacing for the depths of 5 and 10 cm. 
Clearly, the efficiency at first improves as $H$ grows, and then decreases again. But the differences are small. For 10 cm depth, the variations are below 5\%. This, in combination with the significant reduction in local SAR level (36 \% decay for $H=43 mm$ with respect to initial case of $H=18$ mm) indicates that the SAR efficiency ($B_1^{+}/\sqrt{\text{SAR}_{\text{max}}}$) can be increased considerably as is indeed shown by Fig. \ref{in_depth}c. This figure shows that when SAR efficiency is concerned, the optimal antenna-subject spacing has not yet been reached. However, other factors impose additional constraints on the optimal antenna-subject spacing as will be discussed in the next sections. 

\subparagraph*{Inter-element coupling}

\begin{figure}
\center
\includegraphics[width=0.4\linewidth]{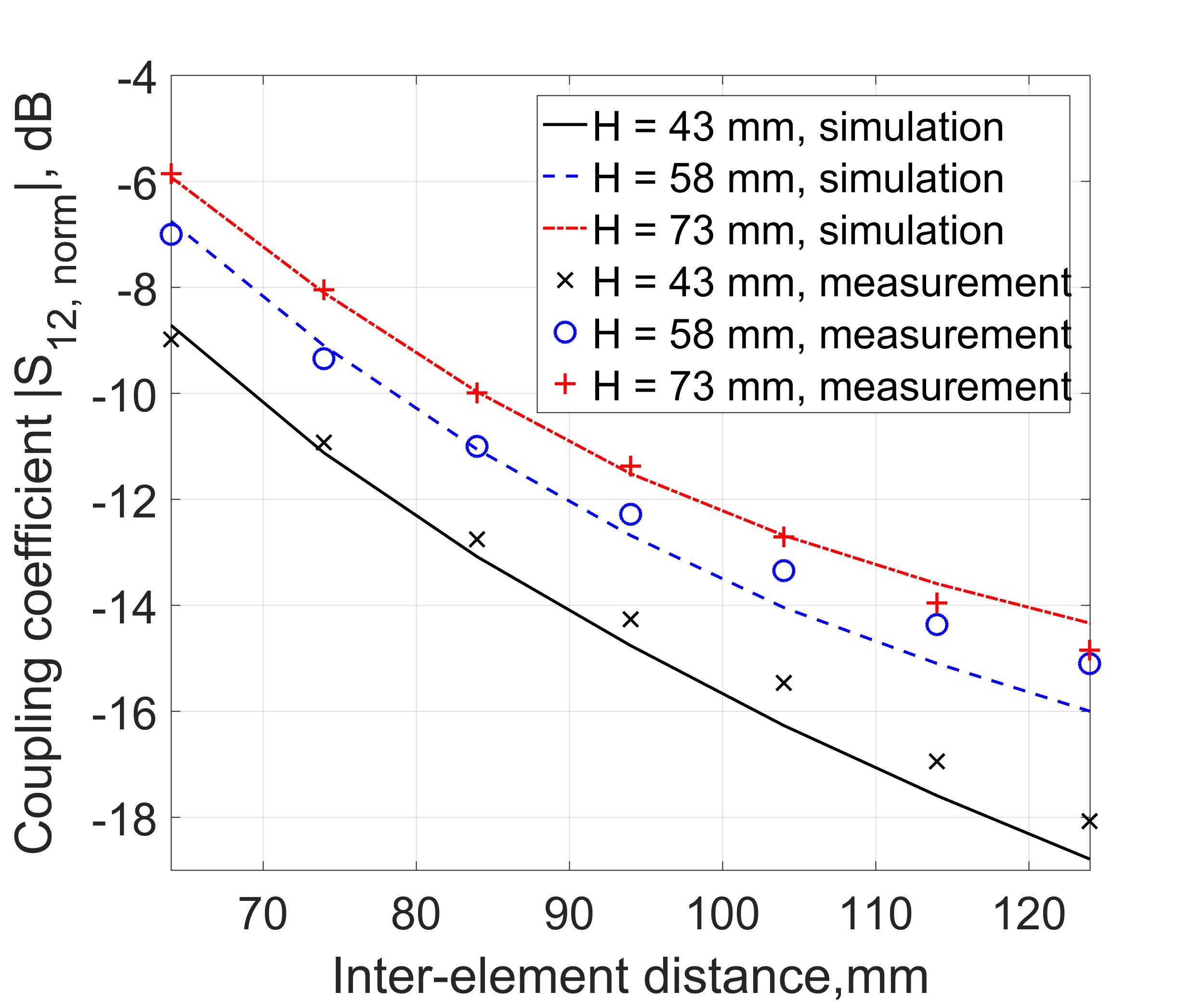}
\caption{Simulated and measured inter-element coupling coefficient vs. inter-element distance $s$ for different antenna-subject spacing $H$}
\label{coupling}
\end{figure}

The inter-element coupling between two fractionated dipole antennas as a function of the inter-element distance $s$ has been determined for the same antenna-subject spacing values as in the previous section by both the on-bench measurements with a phantom and simulations. Fig. \ref{coupling} shows the comparison of the simulated and measured coupling coefficient plotted as a function of the inter-element distance $s$ for $H=$ 43, 58 and 73 mm. 
Simulated and measured data are in very good agreement. Deviations do occur in some graphs for the lowest levels of coupling ($S_{12,\text{norm}}<-13$ dB) but remain below 0.5 dB. 
As expected, coupling between two dipole antennas increases with additional spacing. In general, the coupling is low. When the distance $s$ between the axes of the dipoles is as small as 84 mm, the coupling coefficient ranges from -13 dB for $H=43$ mm to -10 dB for $H=73$ mm therefore staying acceptable for the whole considered range of $H$. Only if the antennas are placed closest together (polycarbonate placeholders are touching each other and $s=64$ mm) the coupling levels become considerable (worse than -10 dB) for all spacings (-9 dB for $H=43$ mm and -5.8 dB for $H=73$ mm). 

\subparagraph*{Robustness towards loading variation}

\begin{figure}[t]
\center
\includegraphics[width=0.8\linewidth]{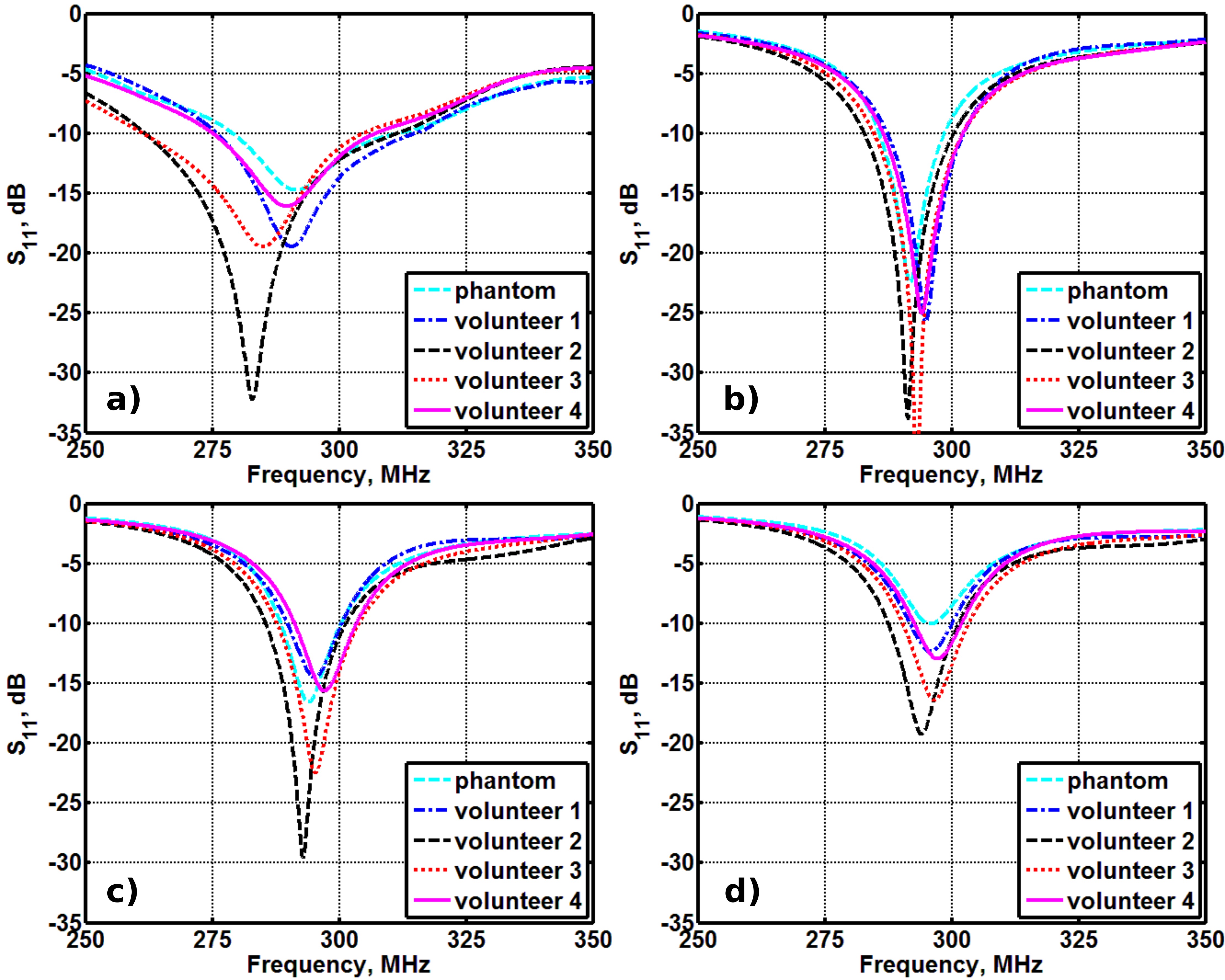}
\caption{Reflection coefficient of a single dipole antenna vs. frequency for different antenna-subject spacing measured in the presence of pelvis-shaped phantom and four volunteers in a dummy bore: a) $H=18$ mm; b) $H=43$ mm; c) $H=58$ mm; d) $H=73$ mm}
\label{load_var}
\end{figure}

The preceding results indicate that for the optimal $B_1^{+}/\sqrt{\text{SAR}_{\text{max}}}$ ratio the thickest spacing is actually the most beneficial in case of the homogeneous phantom. However, practical experience has demonstrated that one fixed matching network is unlikely to provide good matching for a wide range of subjects. To investigate this behavior, the $|S_{11}|$ curve for one antenna with the investigated antenna-subject spacings has been measured on a torso phantom and on four different volunteers. Note that the antenna was schematically tuned and matched in the presence of the phantom for each of the investigated spacings. Then the $S_{11}$ variation was observed when the phantom was replaced by one of four volunteers for each of the investigated spacing values. Results are indicated in Fig. \ref{load_var}.

Looking at all $S_{11}$ curves, it can be seen that the increased sensitivity to loading variations that was expected is not observed. In fact, the largest deviation in the $S_{11}$ curves was obtained for the spacing of 18 mm. At the same time, one can observe that, the reflection at 298 MHz remained below -10 dB for all volunteers with all spacing values. The loaded Q-factor can be roughly estimated by calculating the inverse of the width of the dip in the $S_{11}$ curve. The value obtained increases for increasing spacing from 18 to 43 mm. However, after that the trend is not obvious anymore and in fact it shows a reduction in Q-factor for 58 and 73 mm spacing. We hypothesize that this may be caused by coupling of the antennas to the RF-shield of the bore. For this reason, the antenna-subject spacing for the realized array should not be chosen larger than 43 mm. Visually the $S_{11}$ curve is the most stable for $H=43$ mm, which can be considered the best setup in terms of the inter-subject variations.

\subsection*{Study of 8-channel array configuration}

\subparagraph*{Numerical simulations of 8-channel array}

\begin{figure}
\center
\includegraphics[width=1\linewidth]{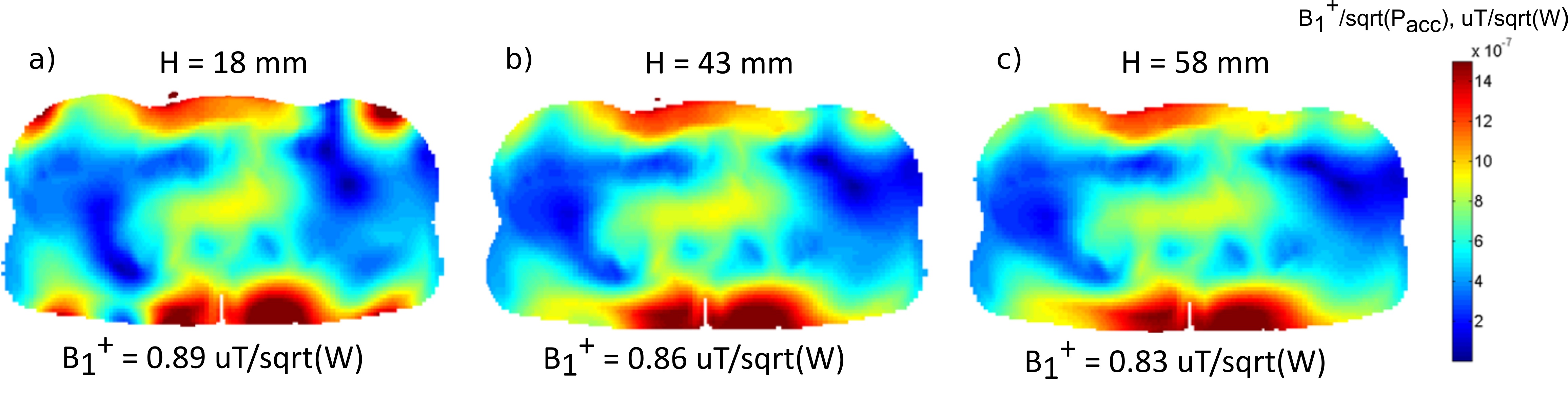}
\caption{Simulation of the 8-element array with different antenna-subject spacing $H$: {$B_1^{+}/{\sqrt{P_{acc}}}$} distribution inside the human model "Duke" in the slice corresponding to the center of the prostate (axial plane) for a) H = 18mm; b) H = 43 mm; c) H = 58 mm}
\label{B1_Duke}
\end{figure}

\begin{figure}[t]
\center
\includegraphics[width=0.95\linewidth]{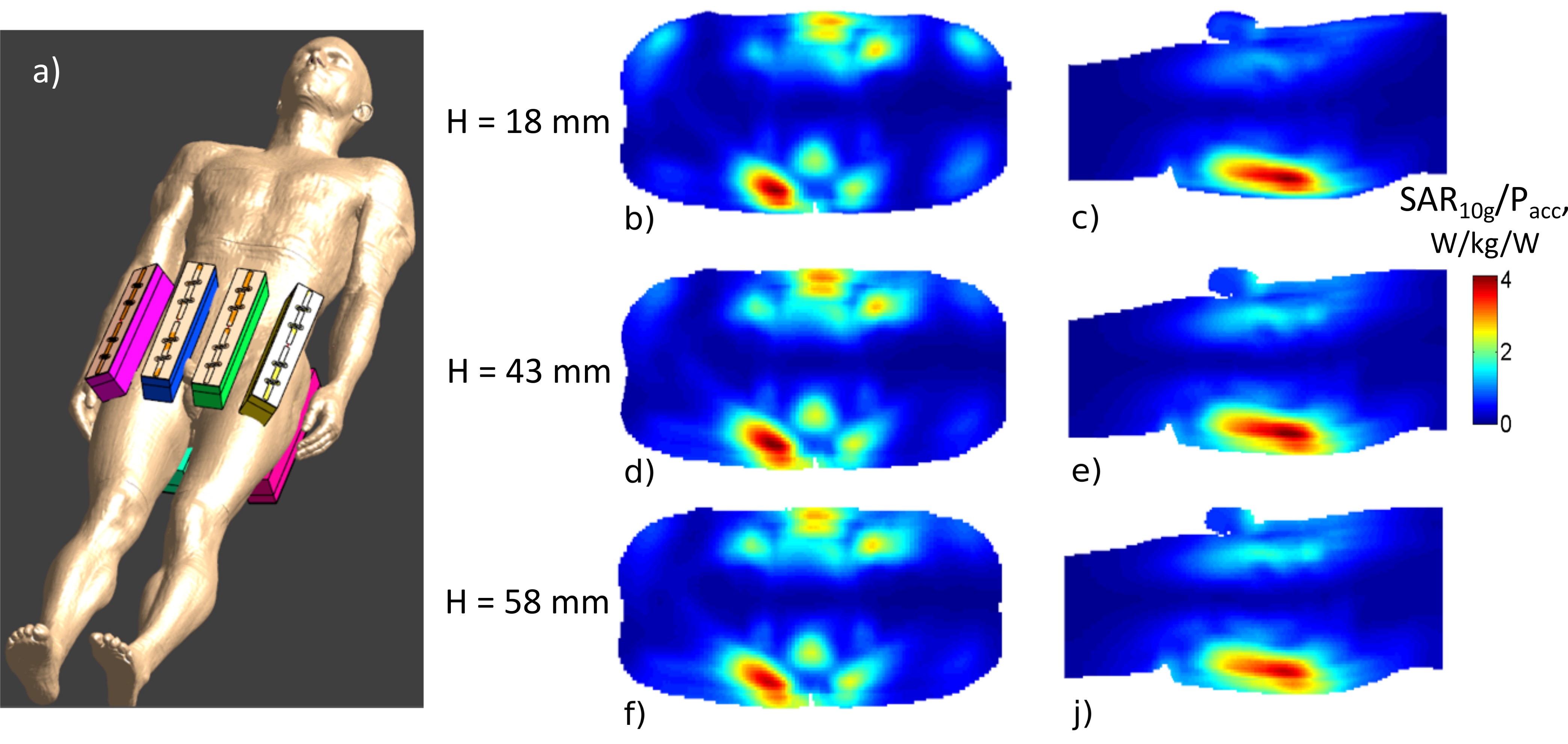}
\caption{Simulation of the 8-element array with different antenna-subject spacing $H$ and the human model "Duke": a) simulation setup (the image was obtained using Sim4Life software); $\text{SAR}_{10g}/P_{acc}$ distribution in the slices corresponding to the maximum value in axial (left column) and sagittal (right column) planes:  b),c) H = 18mm; d),e) H = 43 mm; f),j) H = 58 mm}
\label{SAR_Duke}
\end{figure}

From all presented results obtained using a homogeneous phantom, an optimal antenna-subject spacing of $H=43$ mm has been determined (18 mm polycarbonate + 25 mm foam). The choice was made because of the superior $B_1^{+}/{\sqrt{P_{acc}}}$ efficiency considering that the inter-element coupling is as low as -13 dB for inter-element distance $s=84$ mm typical for the 8-element body arrays. At the same time there is reduction of maximum local SAR in a phantom and there is the best $S_{11}$ curve stability with respect to the subject variation. This distance has been used to realize a 8-element array with increased antenna-subject spacing. The adjacent antennas in the  array configuration (in four antennas on top and in four antennas on the bottom) had the distance $s=84$ mm  as in the original setup with $H=18$ mm \cite{Fractionated}. Simulation results indicate that the $B_1^{+}/{\sqrt{P_{acc}}}$ value inside the prostate slightly decreases with the increase of $H$: 0.89 $uT$ for $H=18$ mm, 0.86 $uT$ for $H=43$ mm, 0.83 $uT$ and for $H=58$ mm (Fig. \ref{B1_Duke}). However, despite that the previous results of SAR simulation obtained with the homogeneous phantom model predicted 36\% reduction of the SAR$/{P_{acc}}$ for a single antenna, the simulation with a realistic body model "Duke" (Fig. \ref{SAR_Duke}a) reveal almost no change of the maximum local SAR (Fig. \ref{SAR_Duke}b-j).

Indeed, as shown in the simulated local SAR plots corresponding to the above discussed $B_1^{+}$ plots, the maximum of the local SAR takes place not at the surface of the subject like in the case of the homogeneous phantom (see Fig. \ref{SAR}) but in the gluteus muscles. Due to a deep location of these muscles the increased antenna-subject spacing does not effect the SAR as it did for the phantom. SAR$/P_{acc}$ value reaches 4.1 $\frac{W}{kg}/W$ for $H=18$ mm, while for the $H=43$ mm it remains almost the same and is equal to 4.2 $\frac{W}{kg}/W$. This can be explained by a high difference in electric conductivity of muscles and fat tissues located nearby a human body surface. The maximum SAR takes place in the most conductive tissues. The maximum local SAR becomes less sensitive to the increase of the antenna-subject spacing when an additional fat layer is located between the muscles and the antenna. As a result in the prostate imaging with both the considered array setups there is almost no difference in $B_1^{+}$ per maximum local SAR when increasing $H$.   

\subparagraph*{$B_1^{+}$ mapping and volunteer imaging}

\begin{figure}[t]
\center
\includegraphics[width=0.9\linewidth]{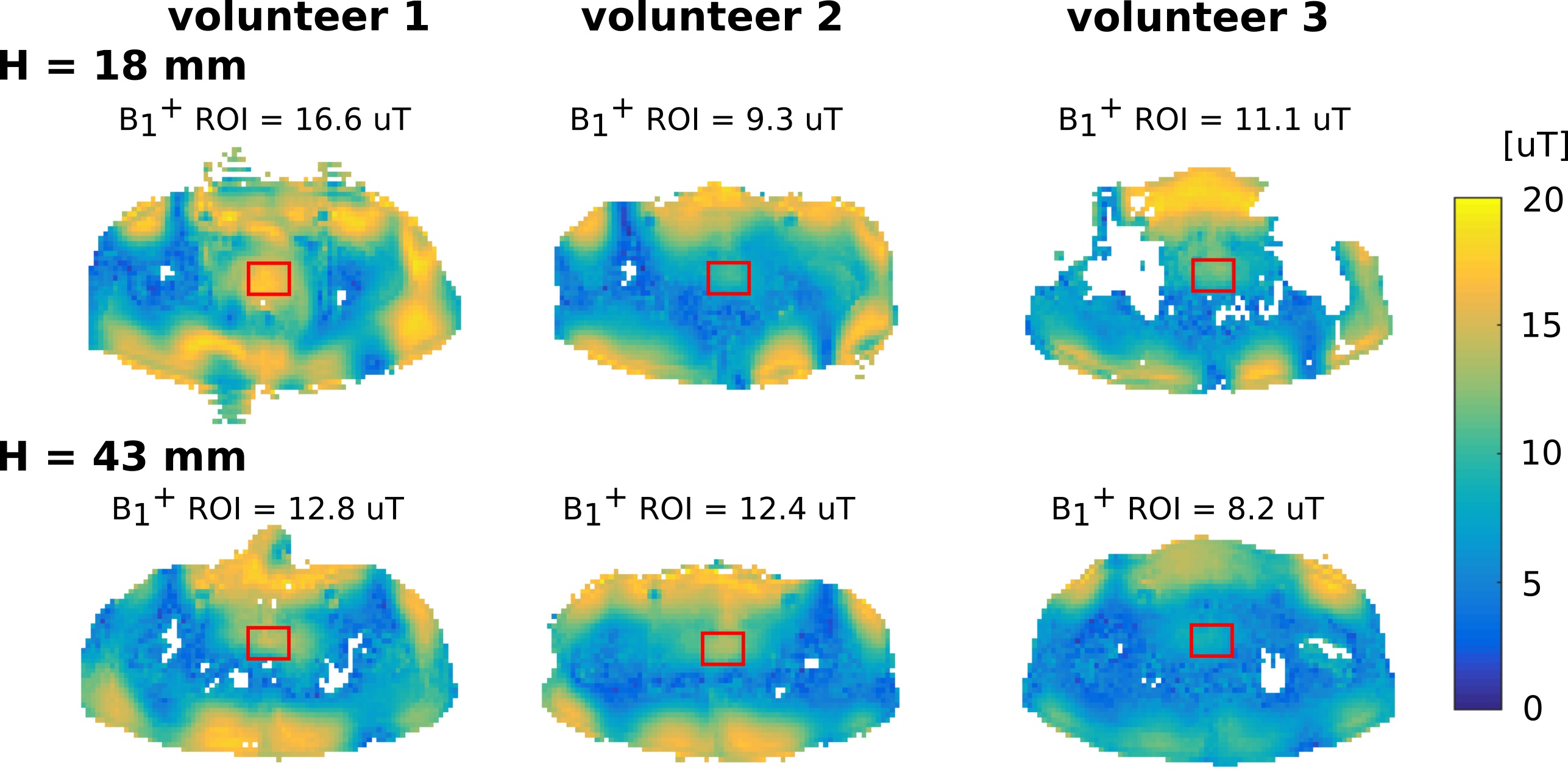}
\caption{$B_1^{+}$ maps of the 8-element array with different antenna-subject spacing $H$ acquired on the 7 Tesla Philips Achieva platform \textit{in-vivo} on 3 volunteers using DREAM sequence for $H=18$ mm (top row) and $H=43$ mm (bottom row)}
\label{3volunteers}
\end{figure}

$B_1^{+}$ maps acquired on 3 volunteers using DREAM sequence for $H=18$ mm and $H=43$ mm are shown in Figure \ref{3volunteers}. Averaged $B_1^{+}$ over 3 volunteers in a selected ROI (prostate region) is slightly lower for H = 43 mm (11.1 uT) as compared to H = 18 mm (12.3 uT) which is in a good correspondence with simulations.
\begin{figure}[ht]
\center
\includegraphics[width=0.9\linewidth]{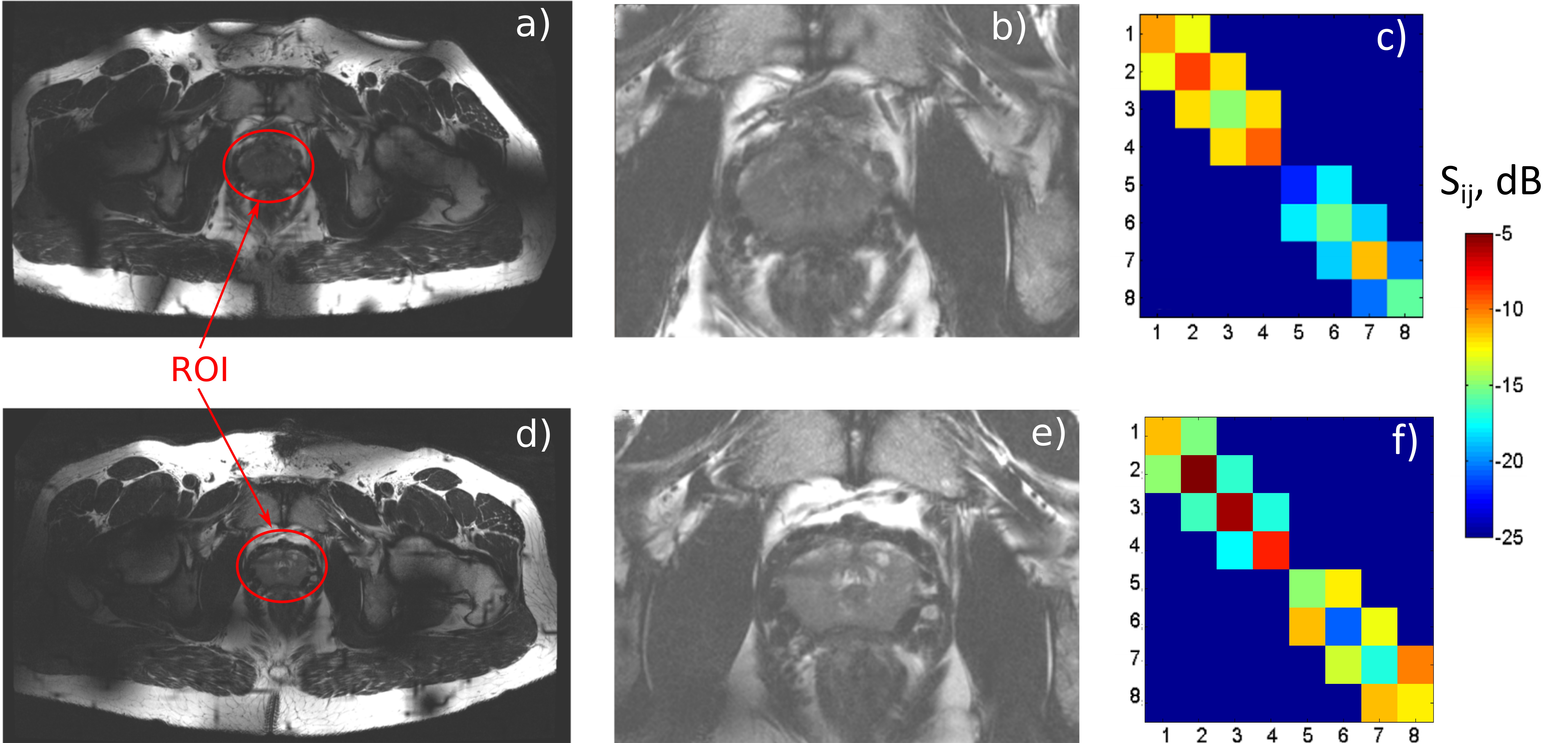}
\caption{\textit{In-vivo} acquired using the 7 Tesla Philips Achieva platform for two 8-element dipole arrays with different antenna-subject spacing $H=18$ mm (first row) and 43 mm (second row): $T_{2}$-weighed images (a,d), detailed image of the prostate region (b,e) and in-bore scattering matrix (c,f)}
\label{T1w+S}
\end{figure}
$T_{2}$-weighed axial-plane images of the whole body acquired on a single volunteer for $H=18$ mm and $H=43$ are indicated in Fig. \ref{T1w+S}a,d while the detailed prostate images are depicted in Fig. \ref{T1w+S}b,e for $H=18$ mm and $H=43$ correspondingly.

The measured S-parameter matrices for the both arrays with $H=18$ mm and 43 mm  indicated in  Fig. \ref{T1w+S}c and f correspondingly. Clearly, the inter-element coupling levels are not dramatic (-12 dB for nearest neighbors, much less for next-nearest neighbors and beyond). 

\section*{Discussion}

This paper provides a general overview of the impact of antenna-subject spacing on peak local SAR level, transmit efficiency, inter-element coupling and sensitivity to loading variations. We hypothesized that SAR levels can be reduced by a larger antenna-subject spacing but that the spacing is constrained by increasing inter-element coupling levels. In fact, coupling turned out to be not very limiting in this perspective and for phantoms, the reduction of SAR levels with increasing antenna-subject spacing was indeed demonstrated for homogeneous phantoms, in contrast to the results observed for simulations on a human model. Indeed SAR level was shown to be not a limiting factor for modifying the antenna-subject spacing of a dipole array for 7T prostate imaging. 

Simulation studies on a phantom have clearly demonstrated the predicted advantage of local SAR reduction. The maximum local SAR level decreases with increasing antenna-subject spacing while the transmit efficiency remains approximately equal. Note that with unchanged transmit efficiency also unchanged receive sensitivity is expected because of the principle of reciprocity (in MRI, the principle of reciprocity states that receive sensitivity is proportional to $B_1^{-}$ while transmit efficiency is proportional to $B_1^{+}$ but the differences in achievable $B_1^{-}$ vs $B_1^{+}$ in this setup and in an array setup are negligible).

In general, the inter-element coupling is really low. At 2 cm antenna-subject spacing and corresponding to the considered case of 8 dipoles around the body, the coupling is -10 dB or less for all investigated antenna-subject spacing values. -10 dB is a considerable amount of coupling but if it comes in exchange for better SAR efficiency it may be worthwhile. Also note that it is only the nearest-neighbor coupling that will reach these values. Results show that next-nearest neighbor coupling will never become an issue for these dipole antennas (when used in a surface coil array at this frequency). For an 8-element array, the inter-element spacing of 20 mm can easily be realized within the circumference of the human body. Even for a 16 element array, this spacing could be feasible, resulting in acceptable inter-element coupling values even if the elements are placed at 76 mm antenna-subject spacing. Only if the antennas are touching each other ($s=$64 mm), the coupling levels will become too high to make up for any potential SAR benefits of the additional spacing. However, all these values are based on measurements and simulations on a phantom with permittivity 81, which is higher than the average tissue permittivity. Therefore, expected coupling values for human loading will be slightly higher. Based on these results on inter-element coupling and assuming an inter-element distance of 2 cm, no limitation is found to increase the antenna-subject spacing to 73 mm. However, the study on the sensitivity towards loading variations provides indications for a different mechanism: coupling of the antennas to the RF shield. This poses the real bottleneck when increasing the antenna-subject spacing. For this reason, the antenna-subject spacing was not increased further than 43 mm (i.e. the value experimentally demonstrated the best inter-subject stability of the $S_{11}$ curve).
Although simulations did not show any change in terms of local SAR level reduction, still a 8-element transmit-receive array was constructed consisting of fractionated dipole antennas with 43 mm antenna-subject spacing. The resulting array has demonstrated almost equal $B_1^{+}$ efficiency in simulations and in $B_1^{+}$ mapping in comparison to the original array with 18 mm antenna-subject spacing and $T_{2}$-weighed images were successfully acquired.

This study has focused on fractionated dipole antennas. However, it is most likely that the qualitative trends are also valid for other types of dipole antennas, although the exact trade-off may be different. For lower field strengths,  (e.g. 3 Tesla) the Q-factor will increase (reduced loading) resulting in enhanced sensitivity to RF shield coupling and/or loading variations. This would suggest a smaller optimal antenna-subject spacing. Vice-versa, higher field strengths with larger frequencies may allow larger antenna-subject spacing with respect to RF shield coupling and/or inter-subject variability.

\section*{Conclusion}

For dipole antennas used in a local transmit or transceive coil arrays for 7T MRI, the effect of the antenna-subject spacing has been investigated. In case of homogeneous phantom study increasing the antenna-subject spacing results in lower peak local SAR levels while the transmit efficiency stays more or less the same or decreases slightly. Overall, the increased spacing enables a higher SAR efficiency ($B_1^{+}/\sqrt(SAR_{max})$) in a phantom. With the increased antenna-subject spacing, the inter-element coupling also increases but this does not pose a problem for an 8-element array at 298 MHz. However, the increase of the antenna-subject spacing makes the antennas more likely to interact with the RF-shield which may severely reduce efficiency. For the design of a single dipole antenna the optimal spacing was found at an additional 25 mm of foam on top of the existing 18 mm of polycarbonate based on the trade-off criteria between the $B_1^{+}/\sqrt(SAR_{max})$ and stability of the $S_{11}$ curve towards subject variations. Moreover, further increase of the antenna-subject spacing would make the array setup inconveniently bulky. With this design, an array of eight dipole antennas has been realized and evaluated for $B_1^{+}$ mapping and volunteer prostate imaging. The obtained results stand in the contrary to the ones obtained using a homogeneous phantom. Simulations did not demonstrate a change in peak local SAR level because the location of peak local SAR was deeply inside the body where the increased thickness of the antennas did not have an impact. The $B_1^{+}$ efficiencies of the original array and the array with increased spacing were similar. Volunteer $B_1^{+}$ mapping confirm that the $B_1^{+}$ efficiency for both arrays remains unchanged. The same can be concluded for the obtained prostate imaging with the two arrays. At the same time the elevated thickness gave better inter-subject stability of $S_{11}$, which means easier tuning and matching of array elements in \textit{in-vivo trials}.

\section*{Acknowledgements}

This project has received funding from the European Union's Horizon 2020 research and innovation program under grant agreement No 736937. This work was supported by the Ministry of Education and Science of the Russian Federation (Zadanie No. 3.2465.2017/4.6). The authors acknowledge Sim4Life by ZMT software (www.zurichmedtech.com).




\end{document}